\documentclass{emulateapj}
\usepackage{apjfonts}

\newcommand{\gtsimeq}{\raisebox{-0.6ex}{$\,\stackrel 
        {\raisebox{-.2ex}{$\textstyle >$}}{\sim}\,$}}

\newcommand{\mgii}{Mg\,{\sc ii}}

\newcommand{\lya}{Ly\,$\alpha$}

\newcommand{\nv}{N\,{\sc v}}

\newcommand{\silii}{Si\,{\sc ii}}

\newcommand{\hi}{H\,{\sc i}}

\newcommand{\myemail}{chris.willott@nrc.ca}

\def\co21{CO\,(2-1)}

\shorttitle{A Lyman alpha halo around a quasar at redshift $z=6.4$}
\shortauthors{Willott et al.}

\begin{document}


\title{A Lyman alpha halo around a quasar at redshift $z=6.4$}


\author{
Chris J. Willott\altaffilmark{1},
Savironi Chet\altaffilmark{1}$^,$\altaffilmark{2},
Jacqueline Bergeron\altaffilmark{3},
John B. Hutchings\altaffilmark{1}
}

\altaffiltext{1}{Herzberg Institute of Astrophysics, National Research Council, 5071 West Saanich Rd, Victoria, BC V9E 2E7, Canada; \myemail}
\altaffiltext{2}{Department of Mathematics \& Statistics, McMaster University, 1280 Main Street West, Hamilton, Ontario L8S 4L8, Canada }
\altaffiltext{3}{Institut d'Astrophysique de Paris, CNRS and Universit\'e Pierre et Marie Curie, 98bis Boulevard Arago, F-75014, Paris, France}

\begin{abstract}

We present long-slit spectroscopic data which reveals extended
\lya\ emission around the $z=6.417$ radio-quiet quasar
CFHQS\,J2329-0301. The \lya\ emission is extended over 15\,kpc and has
a luminosity of $> 8\times 10^{36}$\,W, comparable to the most
luminous \lya\ halos known. The emission has complex kinematics, in
part due to foreground absorption which only partly covers the
extended nebula. The velocity ranges from $-500$\,km\,s$^{-1}$ to
$+500$\,km\,s$^{-1}$, with a peak remarkably close to the systemic
velocity identified by broad \mgii\ emission of the quasar. There is
no evidence for infall or outflow of the halo gas. We speculate that
the \lya\ emission mechanism is recombination after quasar
photo-ionization of gas sitting within a high-mass dark matter
halo. The immense \lya\ luminosity indicates a higher covering factor
of cold gas compared to typical radio-quiet quasars at lower redshift.

\end{abstract}


\section{Introduction}

There are still many mysteries surrounding the cosmological evolution
of the most massive galaxies. Did these galaxies undergo a monolithic
collapse with extremely high star formation efficiency at an early
epoch? Did they all build up supermassive black holes rapidly at their
centers? What processes are responsible for the subsequent halt in
star formation, so that today they appear red and dead?

At high redshift, massive galaxies are extremely rare and are not
found by small area surveys such as with the {\it Hubble Space
  Telescope}. One of the best ways to locate massive galaxies at
$z\approx 6$ is by finding luminous quasars, since these quasars
contain supermassive black holes with masses $\gtsimeq 10^8 M_\odot$
accreting at the Eddington limit (Jiang et al. 2007; Kurk et al. 2007;
Kurk et al. 2009; Willott et al. 2010). The space density of black
holes at this epoch is a factor of $10^4$ times lower than today
(Willott et al. 2010). They are located in rare high matter density
peaks where galaxy formation got underway rapidly (Volonteri \& Rees
2006).

Cosmological simulations show that massive galaxies located within hot
gas halos can accrete a substantial amount of cold gas, so long as
the cold gas is fed into the galaxy in narrow filaments (Dekel et
al. 2009). These cold streams may be the dominant mode of star
formation at high redshift. This is difficult to prove observationally
by \lya\ absorption, due to the small covering factor of the cold streams
(Faucher-Gigu\`ere \& Keres 2011).

Cold gas accretion can also have effects on the \lya\ emission around
galaxies. As cold, neutral gas falls toward the central galaxy it can
be heated and ionized by shocks and recombines emitting \lya\ photons
(Haiman et al. 2000). Cold gas within galaxies could also be ionized
by an intense starburst or an active galactic nucleus (AGN) at the
galaxy center (Haiman \& Rees 2001).  \lya\ ``blobs'' or halos have
been discovered around active and inactive galaxies (e.g. Heckman et
al. 1991; Fynbo et al. 1999; Steidel et al. 2000; Matsuda et al. 2004;
Weidinger et al. 2004; Christensen et al. 2006; Hennawi et
al. 2009). All of the above mechanisms have been proposed to explain
the \lya\ emission from these objects and it is likely that a range of
mechanisms are at work.

CFHQS\,J232908-030158 (hereafter CFHQS\,J2329-0301) is a quasar at
$z=6.417$ discovered in the Canada-France High-z Quasar Survey (CFHQS;
Willott et al. 2007). It is radio-quiet (Wang et al. 2008) and powered
by a black hole with mass $\approx 2 \times 10^8 M_\odot$ (Willott et
al. 2010). There is some evidence for a protocluster of companion
star-forming galaxies based on deep multi-color imaging (Utsumi et
al. 2010). However, most of these objects are visible in the $i'$ band
thumbnail images and therefore likely lie at redshift lower than
$z=6.4$. Goto et al.  (2009; hereafter G09) presented deep broad-band
imaging observations of this quasar and showed that it is spatially
extended at $z'$ band and possibly also at $z_r$ band. The high level
of extended emission at $z'$ band suggests that there is a large flux
due to extended \lya\ emission. Willott et al. (2007) found that the
broad \lya\ line in this quasar has twice the equivalent width of
typical quasars.

We present a deep long-slit spectroscopic observation of
CFHQS\,J2329-0301 with the aim of determining how much of the extended
flux of G09 is due to \lya\ emission and its spatial and kinematic
distributions. In Section 2 we analyze the imaging data of G09. In
Section 3 we describe the spectroscopic observations and their
analysis. Section 4 discusses the physical nature of this galaxy.
All optical and near-IR magnitudes in this paper are on the AB
system. Cosmological parameters of $H_0=70~ {\rm km~s^{-1}~Mpc^{-1}}$,
$\Omega_{\mathrm M}=0.28$ and $\Omega_\Lambda=0.72$ (Komatsu et
al. 2009) are assumed throughout. One arcsecond on the sky corresponds
to a physical size of 5.6\,kpc at $z=6.417$.


\begin{figure*}
\resizebox{1.00\textwidth}{!}{\includegraphics{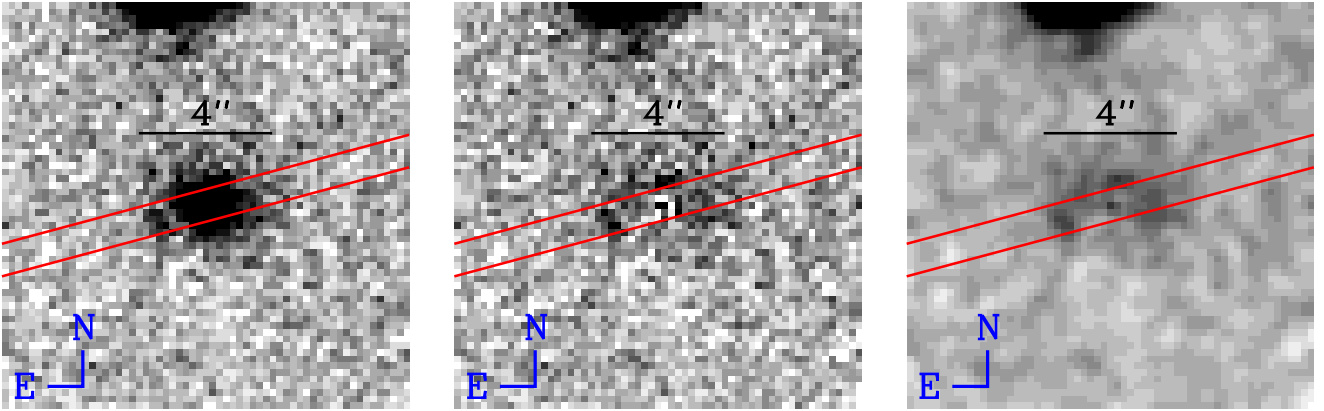}}
\caption{{\it Left:} Subaru Suprime-Cam $z'$ band image of the quasar CFHQS\,J2329-0301 from G09. The red lines marks the location of the Keck ESI 1'' wide spectroscopic slit. {\it Middle:} The same image after subtraction of a model PSF scaled to 0.87 times the total flux. {\it Right:} The PSF-subtracted image after smoothing with a 1 pixel (0.2'') gaussian kernel. All three images are displayed with the same greyscale stretch parameters.
\label{fig:image}
}
\end{figure*}

\section{Extended emission in broad-band images}

G09 presented Subaru Suprime-Cam images of CFHQS\,J2329-0301 in three
broad-band filters: $i'$, $z'$ and $z_r$. The data were obtained in
excellent conditions with seeing of 0.5''. The quasar is very faint in
the $i'$ band ($i'=25.5$) due to IGM absorption shortward of the
\lya\ emission line. Therefore no extended emission in this filter is
observed. However, the data are useful to check that none of the
extended emission found at longer wavelengths is due to a
contaminating object such as a low redshift galaxy. At $z'$ band the
object is clearly more resolved than stars in the field. G09 carried
out point-spread function (PSF) subtraction revealing an elongated
(roughly E-W) object extended over 4 arcsec with magnitude $z'=23.5
\pm 0.3$ in a 2.6'' radius aperture. Due to the lower sensitivity of
Suprime-Cam at $z_r$ band, G09 found only a tentative detection of
extended flux in this filter (which probes longward of the
\lya\ emission line) with magnitude $z_r>24$. The $z'$ band contains
the \lya\ line and the brightness of the extended emission in this
filter suggests that at least some fraction of the extended emission
is from \lya.

In order to compare these imaging observations with our spectroscopy
to be presented in the following section, we have obtained the
processed Suprime-Cam $z'$ band image from the authors of G09. We
performed an independent PSF subtraction using field stars of similar
magnitude to the quasar. Six stars were used to construct a model PSF
which has FWHM 2.4 pixels or 0.48''. The PSF was scaled to match the
total magnitude of the quasar in a circular aperture of radius
2''. This PSF was then subtracted from the quasar image. By trial and
error it was found that the model PSF should be scaled by a factor of
0.87 such that the integrated flux within the central few pixels was
just above zero. Note that this method (maximal PSF-subtraction) therefore gives a lower limit
on the extended flux. We determined an upper limit to the PSF scaling
by considering when the integrated flux within the central few pixels
was more negative than the noise across clean parts of the image. It
was found that the PSF scaling had to be $< 0.89$ at the 4\,$\sigma$
level.
 
The PSF-subtracted residual image (center panel of Figure
\ref{fig:image}) shows a similar structure to that in G09. The
ring-like nature of the residuals could be real or could be an
artifact of the maximal PSF-subtraction. We measure a magnitude of
$z'=23.45 \pm 0.06$ within a 2'' radius aperture, consistent with the
magnitude measured by G09. Note the uncertainty on the magnitude
quoted above is only due to the measured pixel-to-pixel noise in the
sky aperture. It does not include the uncertainty due to PSF scaling
or the effect of correlated pixels. The 4\,$\sigma$ limit based on the
PSF scaling discussed previously corresponds to an extended component
magnitude limit of $z'<23.62$. 

Inspection of the image suggests substantial correlation between
nearby pixels due to a spatially variable background. To ascertain the
significance of the extended emission in the $z'$ band image, we
placed 2'' radius apertures at random across the image (after masking
of real sources) and measured the distribution of fluxes in these
apertures. A flux equivalent to magnitude fainter than $z'=25.2$ is
found in 68\% of the random apertures and magnitude fainter than
$z'=24.6$ is found 95\% of the time. Assuming a Gaussian distribution,
we extrapolate this to infer that the extended $z'$ emission has a
significance of $5\,\sigma$. G09 stated that the extended $z'$
emission has significance $16\,\sigma$. However, their random aperture
results are similar to ours; they found a 1$\,\sigma$ sky noise of
$z'=24.9$, which would make the detection significant at only
$3.6\,\sigma$. In conclusion, the extended emission at $z'$ band is
significant at at least the $5\,\sigma$ level (recalling that we
subtracted off a maximal PSF). With a magnitude of $z'=23.5$ the
extended emission comprises at least 13\% of the total flux from this
object in this filter passband.

\section{Long-slit spectroscopic observations of the extended emission}

\subsection{Observations}

The quasar CFHQS\,J2329-0301 was observed with the Echelle
Spectrograph and Imager (ESI) spectrograph (Sheinis et al. 2002) at
the Keck-II telescope during the nights of 4+5 October 2007 and 24
September 2008. The atmospheric conditions were variable and only the
data taken in reasonable conditions were included. The total
integration time was 8.5 hours, split into 30 minute exposures with
the quasar position offset along the slit between each exposure. The
1'' wide slit was positioned at an angle of 105$^\circ$ East of North,
as indicated on Figure \ref{fig:image}. Note that the spectroscopic
observations were performed before publication of G09, so the
good alignment of the slit with the elongation angle of the extended
emission is coincidental. \lya\ falls in order 7 of the echelle, where
the spatial pixel scale is 0.163'' pixel$^{-1}$ and the spectral scale
is 11.5\,km\,s$^{-1}$ pixel$^{-1}$. The spectral resolution with the
1'' slit is 75 km\,s$^{-1}$ or $R=4000$.

Data reduction was performed using the ESIRedux
code.\footnotemark\ \footnotetext{http://www2.keck.hawaii.edu/inst/esi/ESIRedux/index.html} After
bias removal, a pinhole mask observation was used to determine the
spectral traces of the orders on the detector. The data were then
flat-fielded, sky-subtracted and the wavelength solution determined
using arcs. A standard star observation was used to determine the
target traces. There is significant fringing at the red end of the
spectra, so a master fringe frame was produced by median-combining all
data of faint objects obtained over the two runs with the locations of
known targets masked out. Appropriately scaled fringe frames were
subtracted from each spectrum of the quasar. The individual frames
were then shifted and combined with weighting dependent upon their
signal-to-noise ratios (S/N). Each order was extracted from the 2D
image and rectified using the target trace. The seeing, as measured
from the quasar continuum is 0.93'', which is significantly poorer
than in the imaging observations of G09. Flux calibration is achieved
with standard star spectra and the published broad-band magnitude of
the quasar (Willott et al. 2007).

\begin{figure}
\resizebox{0.48\textwidth}{!}{\includegraphics{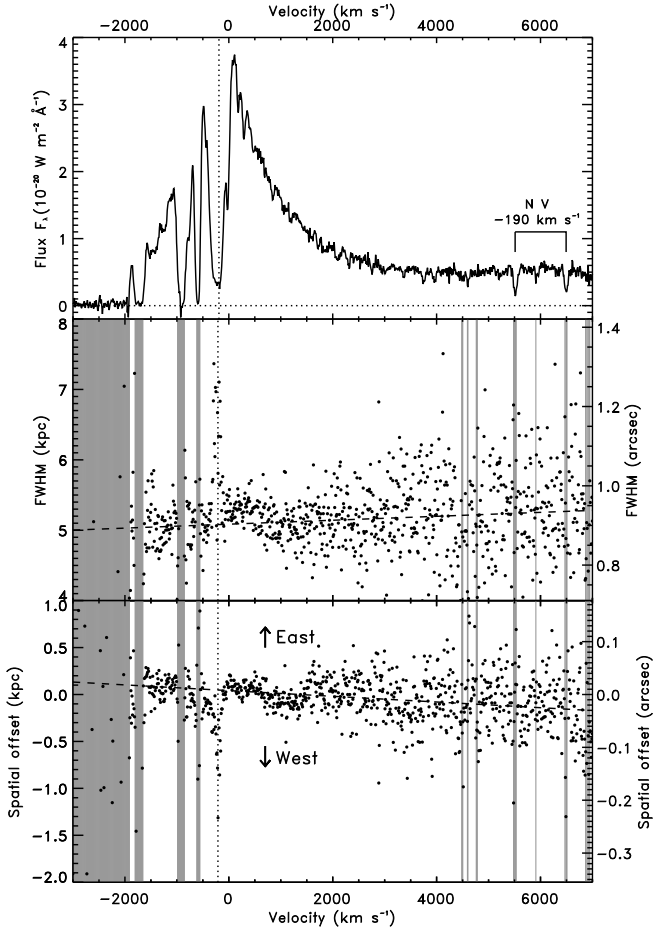}}
\caption{{\it Upper:} Keck ESI extracted spectrum of CFHQS\,J2329-0301. The velocity axis is for the \lya\ line relative to the systemic redshift of $z=6.417$. {\it Middle:} The spatial Gaussian FWHM fitted to each column of the spectrum in units of kpc and arcsec. Regions with S/N\,<\,10 in the extracted spectrum are marked with grey shading, indicating difficulty in fitting the Gaussian. The dashed line show the best fit linear solution to good wavelength regions as described in the text. The dotted line at velocity -190\,km\,s$^{-1}$ marks the location of an absorption feature in the broad \lya\ line with associated \nv\ absorption. At the velocity of this \lya\ absorption feature the flux is more spatially extended than the continuum and broad \lya\ emission. {\it Lower:} Spatial location along the slit of the peak of the fitted Gaussian of each column. The dashed line is the best fit linear solution to good wavelength regions. The spatial centroid of the flux at velocity -190\,km\,s$^{-1}$ is offset by 0.2'' ($\sim 1$\,kpc) from the continuum and broad \lya.
\label{fig:esispecposition}
}
\end{figure}

\begin{figure}
\resizebox{0.48\textwidth}{!}{\includegraphics{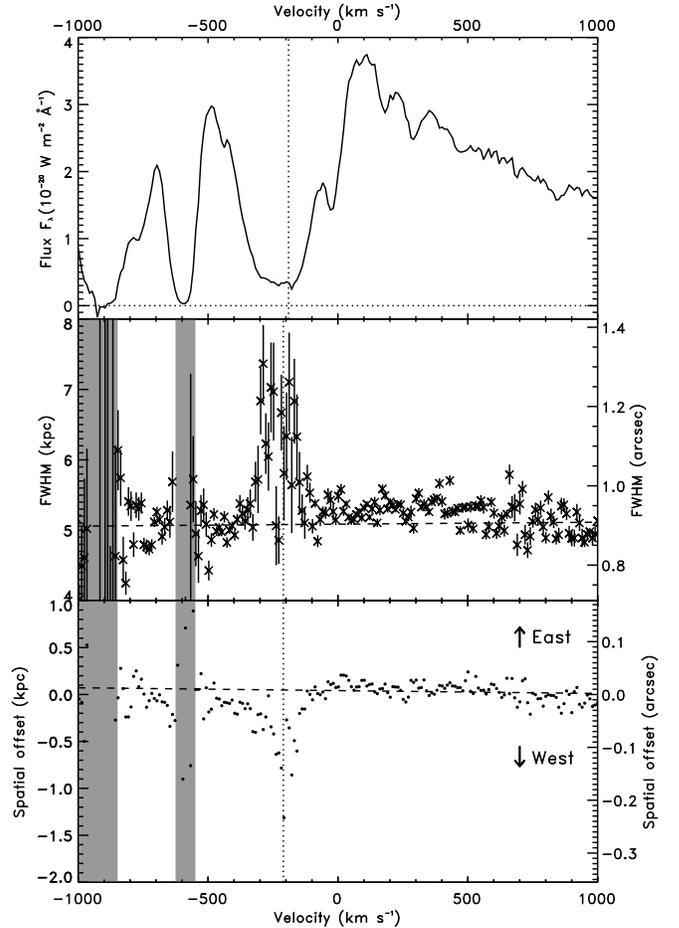}}
\caption{Similar series of plots as in Figure \ref{fig:esispecposition} but focusing on the region around the \lya\ line. Uncertainties on the fitted Gaussian FWHM in the middle panel are plotted. The absorption feature at velocity -190\,km\,s$^{-1}$ is more clearly revealed to have a complex structure. The broad \lya\ line has a FWHM that is about 5\% greater than the linear fit to the continuum.  
\label{fig:esispecpositionzoom}
}
\end{figure}

\subsection{Spectral PSF subtraction}

Due to the relatively short slit of ESI (20'') and the need to dither
the target along the slit, it was not possible to simultaneously
observe a star on the slit to determine the PSF. Therefore the true
PSF for these observations is uncertain. Inspection of the spectrum
showed that there is extended flux at the wavelength of the
\lya\ line. Figures \ref{fig:esispecposition} and
\ref{fig:esispecpositionzoom} show the extracted one-dimensional
spectrum and how the spatial FWHM and peak location along the slit
vary as a function of \lya\ velocity. Zero velocity is set at
$z=6.417$, based on the broad \mgii\ emission line redshift from Willott et
al. (2010). The \mgii\ line is a good estimator of the systemic
velocities of quasars (Richards et al. 2002). 

There are several deep \lya\ absorption lines on the blue wing of the
broad \lya\ line. The broadest, deepest \hi\ absorption dip centred at
$-190$\,km\,s$^{-1}$ ($z_{\rm abs}=6.4123$) is a multi-component
\hi\ system. Fitting the \lya\ profile yields a minimum of three (more
likely four) components, of which one has associated
\nv\ absorption. The total neutral hydrogen column density is $\log
N$(\hi)$\,=14.8\pm 0.2$\,cm$^{-2}$.  The associated \nv\ doublet at
$z_{\rm abs}=6.41241 \pm 0.00005$, marked on Figure
\ref{fig:esispecposition}, is unresolved and highly saturated (both
lines of the doublet have rest equivalent widths $w_{\rm r} = 0.20$
\AA). There is also possible associated \silii\,$1260.4$ absorption at
$z_{\rm abs}=6.4136\pm 0.0001$ but it falls in a region of strong sky
absorption lines.

A Gaussian flux distribution was fitted to each column of the spectrum
to determine the spatial FWHM and peak location (centroid). It was
found that both these quantities were slowly varying functions of
wavelength. Linear fits to both quantities were determined using a
$\chi^2$ minimisation using only wavelength regions with high quality
data and avoiding the obviously more spatially extended parts of the
spectrum. The fitted regions were $-700$ to $-1700$\,km\,s$^{-1}$ and
$+700$ to $+8440$\,km\,s$^{-1}$.

Figures \ref{fig:esispecposition} and \ref{fig:esispecpositionzoom}
reveal two interesting things. Firstly, at the location of the
\hi\ absorption at $-190$\,km\,s$^{-1}$, the \lya\ flux is
significantly more extended than elsewhere with a FWHM of 1.2
arcsec. Secondly, its centroid is significantly offset along the slit
by 0.1'' to 0.2'' ($\approx 1$ pixel) towards the west. Note
that this offset is relatively small compared to the total extent of
extended emission found in the image of G09. Our interpretation of
this is that the absorption system at $-190$\,km\,s$^{-1}$ covers the
quasar nuclear emission more completely than it covers the extended
\lya\ emission. We will return to this point in Section 3.3.

The middle panel of Figure \ref{fig:esispecpositionzoom} shows that
the \lya\ emission at velocity $0$ to $+700$\,km\,s$^{-1}$ has only a
marginally greater FWHM than the continuum fit level marked by a
dashed line. This does not preclude a significant extended component
at this velocity due to the fact that a minor extended component does
not alter the FWHM greatly. For example in the imaging of G09, the
$z'$ band quasar FWHM is equal to the model PSF despite the fact that
13\% of the flux is in the extended component.

In order to determine the PSF we use the Gaussian fit results
described above, which are dominated by data outside of the
\lya\ emission line wavelength range. This makes the very large
assumption that the only extended flux in the spectrum is in the
\lya\ emission line and not in wavelengths corresponding to quasar
continuum. We will return to analyze this assumption in Section 3.3.
Our method is very similar to the spectral PSF (SPSF) method of Moller
(2000; see also Moller et al. 2000). The linear fits to the Gaussian
centroids and FWHM show that these quantities are very slowly varying
functions of wavelength and therefore the values expected at each
wavelength are well-determined from the linear fits.

As a first test of PSF-subtraction we subtracted Gaussians with the
width and centroid determined from the linear fit described above. The
normalization of the Gaussians were set by the observed peak
values. This test showed that there were substantial positive and
negative residuals in the form of wiggles with peaks up to $\pm 4$\%
of the Gaussian peak value. We checked standard star spectra from the
same run and found these wiggles to be present in Gaussians subtracted
off stellar spatial profiles and therefore these wiggles form part of
the instrumental PSF. The wiggles had subtly different form for the
standard stars compared to the quasar data (which was collected over
several different nights) and therefore the stellar spectra could not
be used to define the instrumental PSF. Instead we used the quasar
spectrum at wavelengths equivalent to \lya\ velocities of $+1150$ to
$+2150$\,km\,s$^{-1}$. At this wavelength the quasar is not obviously
extended, yet there is sufficient flux to determine a high S/N PSF
correction to the Gaussian profile. Note that this procedure means
that any true extended flux at this velocity range would be subtracted
off across the full spectrum. We note that our two-step spectral PSF
method is different to the ``look-up table'' approach of Moller
(2000). However, the results for the two methods will be comparable
because both correctly account for the wavelength-dependent PSF
centroid, the wavelength-dependent PSF width and shape and the
pixelization of the data. Evidence for this will be seen in the
cleanliness of the PSF-subtracted quasar continuum (similar to the
very clean PSF-subtracted continuum in Weidinger et al. 2004 using the
look-up table method).

\begin{figure}
\hspace{-0.4cm}
\resizebox{0.50\textwidth}{!}{\includegraphics{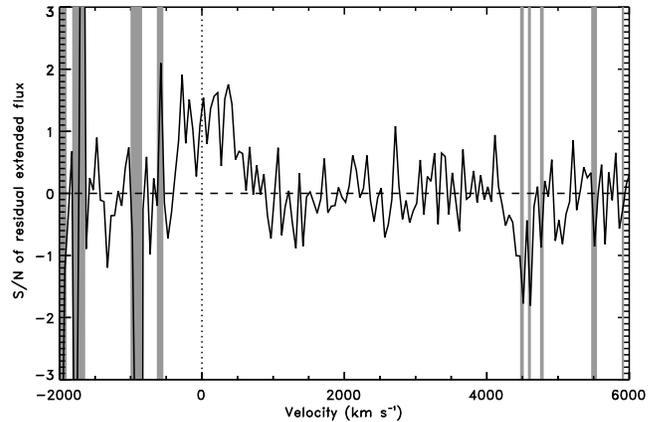}}
\caption{Signal-to-noise ratio of extended flux as a function of wavelength (plotted as \lya\ velocity) for the simple case of PSF-subtraction scaling the PSF to the peak value of each column. The S/N values have been smoothed by 5 pixel averaging. Regions with S/N\,<\,10 in the extracted quasar spectrum are marked with grey shading as in Figures \ref{fig:esispecposition} and \ref{fig:esispecpositionzoom}. Negative S/N residual on this plot indicates negative measured flux due to over-subtraction of the PSF. The most negative regions are those with low S/N in the quasar spectrum where the Gaussian fitting is not reliable. The only large velocity region with positive residual S/N is the $\approx 1000$\,km\,s$^{-1}$ centred on the quasar systemic velocity.  
\label{fig:snrresid}
}
\end{figure}

The final problem is one inherent to PSF-subtraction, how to scale the
PSF in order to just remove the unresolved component without
over-subtracting the extended emission. By simply scaling the PSF peak
to the observed peak, the center of the recovered extended emission
inevitably has zero flux. The first step in this process is to
determine at what wavelengths we have unequivocal evidence for
extended emission. This was achieved by performing spectral
PSF-subtraction with the PSF peak set equal to the observed peak such
that there is zero flux in the extended component at the center. The
residual flux from this process was then integrated along each column
to determine the total extended flux per wavelength pixel. This total
residual flux was then divided by the total noise in each column of
the spectrum, taking into account sky and object contributions to the
noise. This resulted in the S/N of extended emission as a function of
wavelength and is plotted in Figure \ref{fig:snrresid}. The only
velocity region which shows positive extended flux residual over a
sizeable velocity range is between $-400$ and
$+470$\,km\,s$^{-1}$. Even though the S/N of each column is only
between 1 and 2, the integrated S/N across this velocity range is 10
and highly significant.

Recall that the velocity region $+1150$ to $+2150$\,km\,s$^{-1}$ was
used to determine the non-Gaussian component of the PSF, so any
extended flux at that velocity would be subtracted off as part of the
PSF. Whilst we cannot conclude that there is no extended emission at
$> +500$\,km\,s$^{-1}$ (see also Section 3.3), we can say that there is
a greater amount of extended emission located close to the quasar
\lya\ center than at longer wavelengths.

Therefore the PSF is scaled to the observed peak outside of the
velocity region $-400$ and $+470$\,km\,s$^{-1}$. Within this region
the PSF scaling is determined using the observed peak, but with a
correction if necessary to prevent the total flux of the extended
component within the central 2 or 3 spatial pixels being
negative. Note that this is still a maximal subtraction and it is
possible that the extended component contains a greater fraction of
the total flux.

\subsection{Results}

\begin{figure}
\vspace{0.7cm}
\hspace{0.6cm}
\resizebox{0.44\textwidth}{!}{\rotatebox{90}{\includegraphics{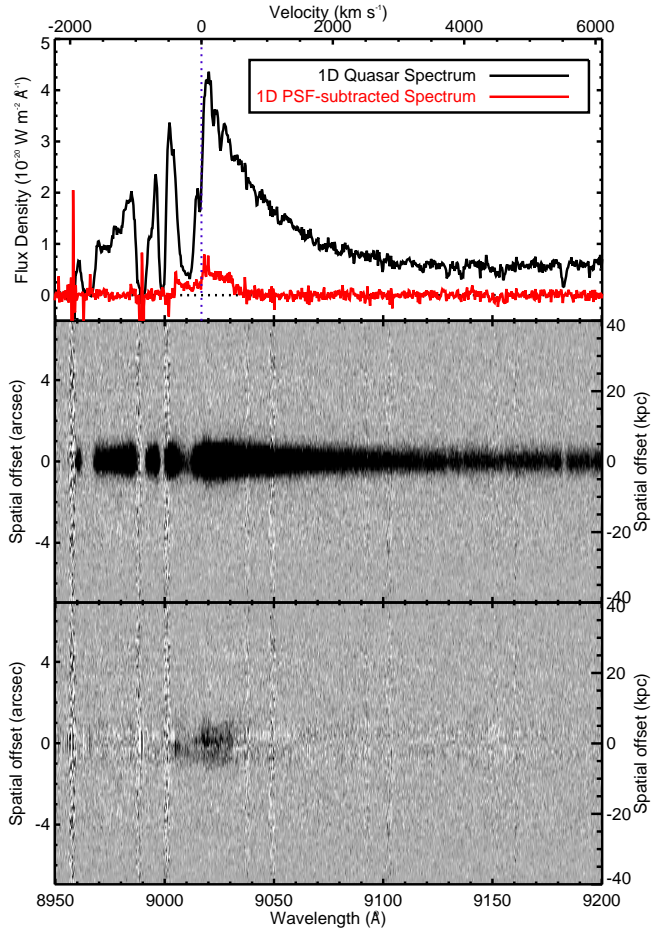}}}
\caption{{\it Upper:} One-dimensional extracted quasar spectrum (black) compared to the PSF-subtracted residual spectrum of the extended emission (red). {\it Middle:} Original 2D spectrum. {\it Lower:} PSF-subtracted 2D spectrum showing the extended \lya\ halo. Positive spatial offset corresponds to the east and negative to the west.
\label{fig:twodspec}
}
\end{figure}

The results of the spectral PSF-subtraction process are shown in
Figure \ref{fig:twodspec}. The upper panel compares the quasar
spectrum with the spectrum of the extended component. The middle panel
shows the original 2D spectral image and the lower panel the 2D
spectral image extended residuals after PSF-subtraction. There are
very few positive or negative residuals at velocities $<-400$ or
$>+500$\,km\,s$^{-1}$ which shows how well our spectral PSF-modeling
works. The extended emission appears clumpy both spatially and
kinematically. This structure will be investigated further below.

One of the surprising things about the residual extended emission is
that its flux is so low. Recall that in Section 2 we showed that the
extended emission identified in Subaru imaging contributes 13\% of the
total $z'$ band flux, assuming a maximal PSF-subtraction. To compare
to the extended flux in the spectrum we need to consider differential
slit losses for the PSF and extended emission and the wavelength range
over which to integrate. Differential slit losses for the ESI seeing
based on the Subaru imaging show that we expect 10\% of the total
observed spectroscopic $z'$ band flux to come from extended emission.

Direct measurement of the flux between 9000 and 9040\,\AA\ in the
upper panel of Figure \ref{fig:twodspec} shows that the extended
emission contributes 10\% of the total flux over this narrow
wavelength range containing the extended \lya\ emission. Over a wider
wavelength range of 9000 to 9100\,\AA, the extended component is only
5\% of the total flux.  Based on the Suprime-Cam $z'$ band response
curve, 39\% of the flux from the quasar in this filter comes from 9000
to 9100\,\AA. Therefore, only 2\% of the $z'$ band flux in the ESI
slit comes from the extended component, compared to the expected
10\%. The majority of the extended emission observed in Subaru imaging
is unaccounted for in our spectroscopy.

Possible reasons for this missing extended flux are that the
broad-line region/continuum flux (which we used to determine the PSF)
is also spatially resolved or that we have substantially overestimated
the PSF scaling for the narrow part of the \lya\ line. G09 proposed
that 40\% of the extended $z'$ band emission was continuum flux from
an extreme starburst ($M_{1450}=23.9$) in the host galaxy, based on
their tentative detection of extended emission at $z_r$ band. We have
no evidence for extended continuum based on Figure
\ref{fig:snrresid}. Also the PSF-fit appears to do an equally good job
at accounting for all the observed ESI flux at velocities
corresponding to the broad \lya\ emission (velocities $-2000$ to
$-500$\,km\,s$^{-1}$ and $+500$ to $+2000$\,km\,s$^{-1}$) as it does
at longer wavelengths where the quasar continuum dominates. Even the
extreme starburst advocated by G09 would be a minor component in the
broad \lya\ wavelength region, so this argues against significant
extended continuum. It is more likely that we have overestimated the
PSF scaling for the \lya\ line, since we have no independent
constraint on this normalization. In this case up to half of the
narrow \lya\ emission could be in the extended component. The fact
that our seeing is substantially worse than that of the imaging
observations is another complicating factor.

\begin{figure}
\vspace{0.1cm}
\hspace{-0.3cm}
\resizebox{0.48\textwidth}{!}{\includegraphics{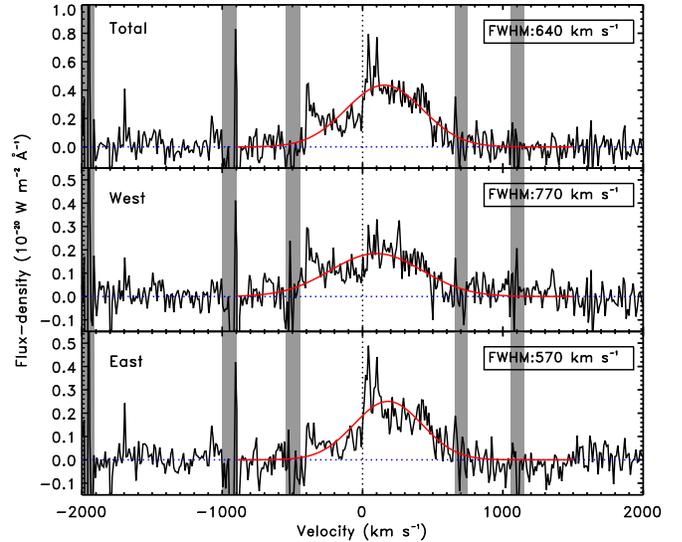}}
\caption{\lya\ halo flux velocity profile for the eastern component ({\it lower}), the western component ({\it middle}) and the total ({\it upper}). Wavelength regions with S/N\,<\,10 in the extracted 1D quasar spectrum are marked with grey shading as in Figures \ref{fig:esispecposition} and \ref{fig:esispecpositionzoom}.  The red curves show Gaussian fits to the velocity profiles. None of the velocity profiles are well fit by a Gaussian.
\label{fig:velprofile}
}
\end{figure}

\begin{figure}
\vspace{0.1cm}
\hspace{-0.3cm}
\resizebox{0.48\textwidth}{!}{\includegraphics{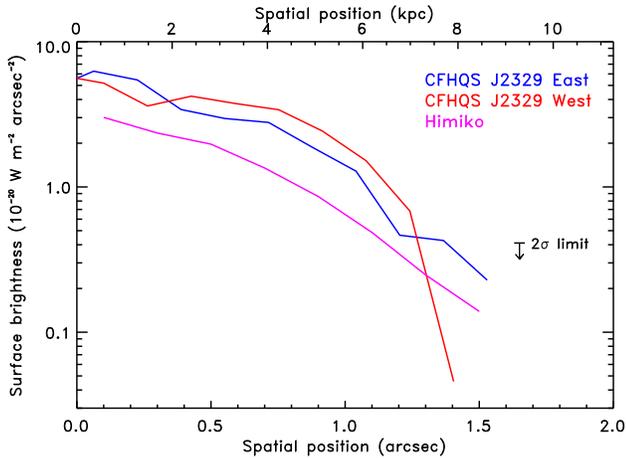}}
\caption{Radial profile of surface brightness for the \lya\ halo of
  CFHQS\,J2329-0301 based on ESI spectroscopy. Different curves are
  plotted for the east and west components. Also plotted is the
  surface brightness profile from narrow-band imaging (filter
  NB921) of the $z=6.6$ \lya\ blob Himiko of Ouchi et al. (2009). The
  2\,$\sigma$ limit arrow shows the 2\,$\sigma$ surface brightness
  limit for the CFHQS\,J2329-0301 data. All curves are plotted out to a
  distance of one data point beyond the last 2\,$\sigma$ detection.
\label{fig:sbprofile}
}
\end{figure}

Even though our observations leave some puzzle about the bulk of the
extended emission, it is interesting to consider the spatial structure
and kinematics of the resolved \lya\ emission. Figure
\ref{fig:velprofile} shows the velocity profile for the whole
\lya\ halo and for the two spatially distinct components on the west
and east sides. The overall velocity structure for the two sides is
quite similar. The complex structure means that neither component is
well fit by the Gaussians plotted in Figure \ref{fig:velprofile}. The
FWHM of these Gaussians do however give a good indication of the
velocity range of the bulk of the \lya\ emitting gas. Both contain a
pair of narrow peaks just redward of zero velocity (more prominent in
the east component).

Both the east and west show a broad dip at $-190$\,km\,s$^{-1}$ due to
partial covering by the $z_{\rm abs}=6.412$ \lya\ absorption system
discussed in Section 3.2. In the eastern component this dip reaches to
zero flux at $-150$\,km\,s$^{-1}$. The eastern component is more
strongly absorbed than the west, consistent with the spatial offset
marked in the lower panels of Figures \ref{fig:esispecposition} and
\ref{fig:esispecpositionzoom}. To constrain further the difference in
spatial coverage of the absorber, we estimated the velocity spread
$\delta v$ of the absorption in the quasar and in the total extended
emission spectra. At an optical depth $\tau=0.1$, $\delta v =
500$\,km\,s$^{-1}$ towards the quasar, whereas it is only
$360$\,km\,s$^{-1}$ for the halo, the latter not extending as far in
the blue as that towards the nuclear emission. Using the four
component fit of the \lya\ absorption profile, we find a velocity
spread $\delta v = 435$\,km\,s$^{-1}$ for the two outermost
components. The total rest equivalent widths of \lya\ in the quasar
and extended spectra also roughly constrain the variations in the
amount of gas in the $z_{\rm abs}=6.412$ absorber. We find $w_{\rm r}
= 1.31 \pm 0.04$ and $0.72 \pm 0.07$ \AA\ for the quasar and halo,
respectively. This suggests a significant difference in the amount of
absorbing gas towards the nucleus and the extended halo.

Figure \ref{fig:sbprofile} shows the surface brightness profile along
the slit in each direction from the quasar. Emission on both sides of
the slit is visible up to 1.5 arcsec from the quasar, similar to the
physical extent visible in Subaru imaging (Figure \ref{fig:image}).
The two components have similar surface brightness distributions,
although the western component appears to have a sharp cutoff at
distance 1.3 arcsec.

 Also plotted on Figure \ref{fig:sbprofile} is the radial profile of
 the $z=6.6$ \lya\ blob Himiko from the narrow-band imaging of
 Ouchi et al. (2009). The size and surface brightness profiles of
 these two objects are very similar, although the halo of
 CFHQS\,J2329-0301 is a factor of 2 brighter than Himiko. This
 will be discussed further in Section 4.

\section{Discussion}

The \lya\ halo around CFHQS\,J2329-0301 has a physical extent of at
least $15$\,kpc. By comparison, typical star-forming galaxies at $z=6$
have a half-light radius of $1$\,kpc (Bouwens et al. 2004). The
\lya\ halo contains gas with a velocity range of $\approx
1000$\,km\,s$^{-1}$. The total emission line flux within the 1 arcsec
ESI slit is $8 \times 10^{-20}$\,W\,m$^{-2}$ which gives a line
luminosity of $4 \times 10^{36}$\,W. Accounting for the fact that only
half the flux identified in the Subaru imaging of Section 2 is located
within the ESI slit and that the majority of the extended emission
from imaging is not identified in our spectroscopy, the total flux and
luminosity are at least twice as great as this, i.e. $> 8\times
10^{36}$\,W.

The halo line luminosity is comparable to the most luminous
\lya\ blobs known (Steidel et al. 2000). Christensen et al. (2006)
showed that radio-quiet quasars at $3<z<4$ have \lya\ halos with
luminosities typically 0.5\% of the broad \lya\ luminosity. By
comparison, for radio-loud quasars (Heckman et al. 1991), this ratio
is typically 5\%. For CFHQS\,J2329-0301 the ratio is 5\%, despite the
fact it is not a strong radio source (Wang et al. 2008).

Interpreting the physical nature of \lya\ halos is notoriously
difficult. \lya\ is resonantly scattered so the observed spatial and
kinematical distributions are not those where the photons were
emitted. \lya\ photons can be generated from several different
mechanisms including shock heating of infalling/outflowing gas or
photo-ionization by stars or an AGN. For the case of
CFHQS\,J2329-0301, we know that the center of the galaxy harbors a $2
\times 10^8 M_\odot$ black hole accreting at the Eddington limit
(Willott et al. 2010) which is generating a lot of \lya\ emission
within the quasar broad-line region (upper panel of Figure 2). The
influence of the quasar UV emission extends 2.5\,Mpc along our
line-of-sight as evidenced by the transmitted flux at
$-2000$\,km\,s$^{-1}$ in the upper panel of Figure 2 (see also Willott
et al. 2007). This makes the AGN explanation to power the \lya\ halo
a distinct possibility.  The size and surface brightness of the
observed halo is within the range predicted by Haiman \& Rees (2001)
for cold gas photoionized by a quasar. This implies the galaxy is at
an early stage in its evolution due to the copious amount of cold gas
located at large distances from the nucleus. The higher ratio of halo
to broad-line luminosity compared to radio-quiet quasars at $3<z<4$
could be due to a greater covering factor of cold gas at higher
redshifts, but a larger sample at $z\approx6$ is required.

What clues come from the gas kinematics? The gas is approximately
centred at the quasar systemic velocity when accounting for the
absorption by the system at $-190$\,km\,s$^{-1}$. The infall models of
Villar-Martin et al. (2007) and Weidinger et al. (2004) both expect a
spatial offset of the peak emission, a velocity offset up to
$500$\,km\,s$^{-1}$ and a velocity FWHM that varies systematically
across the object. None of these features are observed for this
\lya\ halo. Therefore there is no evidence that the gas is dominated
by either outflowing or infalling gas.

The velocity FWHM of $640$\,km\,s$^{-1}$ reflects the velocity range
of the gas where the \lya\ photons are emitted and scattered within
the central 15\,kpc diameter of the host galaxy. This FWHM is much
higher than that of the infalling \lya\ gas in the $z=3$ quasar
Q\,1205-30 (Weidinger et al. 2004) or of Himiko which has
FWHM$=250$\,km\,s$^{-1}$. However, it is comparable to other
\lya\ blobs found at $z=3$ which are suggested to be the sites of
massive galaxy formation based on these large line widths (Matsuda et
al. 2006) and to the quasar \lya\ halos observed by Christensen et
al. (2006). 

Making the possibly unreasonable assumption that the gas is in virial
equilibrium at a typical distance of 7\,kpc with velocity dispersion
$\sigma=640/2.35=270$\,km\,s$^{-1}$, one can determine a dynamical
mass of $M=6\times 10^{11} M_\odot$ using $M=(5/3)(3\sigma^2)R/G$.
Haiman \& Rees (2001) show that even massive halos such as that based
on this dynamical mass can contain substantial quantities of cold gas
in their inner regions due to the short cooling time. We note in
passing that elliptical galaxies in the local universe with
$\sigma=270$\,km\,s$^{-1}$ typically contain black holes with the same
mass as CFHQS\,J2329-0301 (Tremaine et al. 2002). This is in contrast
to the results of Wang et al. (2010) which found SDSS $z\approx 6$
quasars have black hole mass to bulge mass ratios a factor of $\approx
15$ greater than locally.

\section{Conclusions}

We have spectroscopically identified a large and extremely luminous
\lya\ halo around one of the most distant known quasars. The
\lya\ properties do not show signatures of either infall or
outflow. The power source for the \lya\ emission is likely ionization
of cold neutral gas by the strong UV continuum of the quasar. A
neutral hydrogen and highly-ionized metal absorption system in the
foreground is more strongly absorbed towards the quasar nucleus and
eastern component of the \lya\ halo than towards the western component.

The \lya\ luminosity of the halo relative to the broad line region is
at least ten times greater than for typical quasars at redshift
$z\approx 3$ (Christensen et al. 2006). This could indicate a greater
covering factor of cold neutral gas at higher redshifts. However, this
is the only quasar at $z>6$ which has been observed to have extended
\lya\ emission. It will be important to study other high-redshift
quasars to provide a statistical measure of the evolution in the
properties of \lya\ halos. Narrow-band imaging centred on the narrow
\lya\ line wavelength and integral-field spectroscopy will be the most
important observational tools to examine this issue.

\acknowledgments Thanks to Yousuke Utsumi and Tomo Goto for providing
the Subaru Suprime-Cam imaging data and to Masami Ouchi for the
surface brightness profile of Himiko. Thanks to the referee for useful
suggestions that improved the paper. The data presented herein were
obtained at the W.M. Keck Observatory, which is operated as a
scientific partnership among the California Institute of Technology,
the University of California and the National Aeronautics and Space
Administration. The Observatory was made possible by the generous
financial support of the W.M. Keck Foundation.

\end{document}